\documentclass[onecolumn, floatfix, prd]{revtex4-2}
\usepackage{natbib}
\usepackage{silence}
\usepackage{lmodern} 
\usepackage[T1]{fontenc}
\usepackage{textcomp}
\usepackage{xcolor}
\usepackage[scr=boondoxo,frak=boondox,bb=boondox]{mathalfa}
\pdfoutput=1
\usepackage{makecell, multirow}
\setcellgapes{3pt}
\usepackage{hyperref}
\usepackage{amsmath}
\usepackage{tabularx}
\usepackage[vlines]{tabularht}
\usepackage{adjustbox}
\usepackage{tikz}
\usepackage{draftfigure}
\usepackage{amssymb}
\usepackage{graphicx}
\usepackage{float} 
\usepackage{color}
\usepackage{color,soul}
\usepackage{ulem}
\usepackage{multirow}
\usepackage{epsfig}
\usepackage{epstopdf}
\usepackage{rotating}
\usepackage{afterpage}
\usepackage{soul}
\usepackage{fancyvrb}
\usepackage{lineno}
\usepackage{amsmath}
\usepackage{xspace}

\usepackage{amssymb,latexsym,mathrsfs}
\usepackage{multirow}
\usepackage{bigints}
\usepackage{soul}
\usepackage{graphicx}
\usepackage{subcaption}
\usepackage{dcolumn}
\usepackage{bm}
\usepackage{epsfig}
\usepackage{caption}
\usepackage{color}
\usepackage{multirow}
\usepackage{makecell}
\usepackage{textcomp}
\usepackage[title]{appendix}
\usepackage{tabularx}

\newcommand{\be}{\begin{equation}}
\newcommand{\ee}{\end{equation}}
\newcommand{\bea}{\begin{eqnarray}}
\newcommand{\eea}{\end{eqnarray}}
 
\definecolor{mediumpurple}{rgb}{0.58, 0.44, 0.86}

\definecolor{owngreen}{rgb}{0.0, 0.5, 0.0}

\usepackage{hyperref}
\usepackage{url}
\hypersetup{
    colorlinks=true,
    linkcolor=red,
    citecolor=blue,
}

\begin{document} 
\title{Exploring Alternative Cosmologies with the LSST: Simulated Forecasts and Current Observational Constraints}

\newcommand{\URLSNANA}{\url{https://github.com/RickKessler/SNANA}}
\newcommand{\URLLSST}{\url{www.lsst.org}}
\newcommand{\URLDESC}{\url{https://lsstdesc.org}}
\newcommand{\URLOPSIM}{\url{http://opsim.lsst.org/runs/minion_1016/data/minion_1016_sqlite.db.gz}}
\newcommand{\HOSTLIB}{{\tt HOSTLIB}}
\newcommand{\SALTII}{{\sc SALT-II}}
\newcommand{\DESSN}{DES-SN}
\newcommand{\lowz}{low-$z$~}
\newcommand{\mosfit}{{\tt MOSFiT}\xspace}
\newcommand{\bands}{$ugrizy$}
\newcommand{\PLASTICC}{Photometric LSST Astronomical Time Series Classification Challenge}
\newcommand{\acro}{{\tt PLAsTiCC}}
\newcommand{\SNANA}{{\tt SNANA}}
\newcommand{\pippin}{{\tt Pippin}}
\newcommand{\LSST}{Large Synoptic Survey Telescope}
\newcommand{\DES}{Dark Energy Survey}
\newcommand{\OPSIM}{\verb|OpSim|}
\newcommand{\Spec}{Spectroscopic}
\newcommand{\spec}{spectroscopic}
\newcommand{\specy}{spectroscopically}
\newcommand{\ZPHOT}{{\bf\tt ZPHOT}}
\newcommand{\zSpec}{z_{\rm spec}}
\newcommand{\LCDM}{\Lambda{\rm CDM}}
\newcommand{\OL}{\Omega_{\Lambda}}
\newcommand{\OM}{\Omega_{\rm M}}
\newcommand{\zcmb}{z_{\rm cmb,true}}
\newcommand{\RateUnit}{{\rm yr}^{-1}{\rm Mpc}^{-3}}
\newcommand{\Ftrue}{F_{\rm true}}
\newcommand{\sigF}{\sigma_{F}}
\newcommand{\sigFi}{\sigma_{F,i}}
\newcommand{\sigFtrue}{\sigma_{\rm Ftrue}}
\newcommand{\zSN}{z_{\rm SN}}
\newcommand{\zHOST}{z_{\rm HOST}}
\newcommand{\Ngen}{$N_{\rm gen}$} %

\newcommand{\FiData}{F_i^{\rm data}}
\newcommand{\FiModel}{F_i^{\rm model}}
\newcommand{\xvecSALT}{\vec{x}_5}

\newcommand{\sigFiTilde}{\Tilde{\sigma}_{F,i}}
\newcommand{\logsigmodel}{2\ln({\sigFi/\sigFiTilde})}
\newcommand{\sigzhost}{\sigma_{z,{\rm host}}}

\newcommand{\NSAMPLE}{25} 
\newcommand{\NBIASCORTOT}{$3.1 \times 10^6$}
\newcommand{\NBIASCORHIZ}{$2.6 \times 10^6$}
\newcommand{\NBIASCORLOZ}{$4.4 \times 10^5$}
\newcommand{\NSYST}{7}
\newcommand{\dz}{\delta z}
\newcommand{\fout}{f_{\rm out}}
\newcommand{\sigIQR}{\sigma_{\rm IQR}}
\newcommand{\ztrue}{z_{\rm true}}
\newcommand{\zphot}{z_{\rm phot}}
\newcommand{\zspec}{z_{\rm spec}}
\newcommand{\zcheat}{z_{\rm cheat}}
\newcommand{\zhost}{z_{\rm host}}
\newcommand{\Dz}{\Delta z_{(1+z)}}
\newcommand{\Pfit}{P_{\rm fit}}

\newcommand{\G}{G_{\rm host}}
\newcommand{\mubias}{\Delta\mu_{\rm bias}}
\newcommand{\mutrue}{\mu_{\rm true}}

\newcommand{\dzsyst}{\Delta z_{\rm syst-z}}
\newcommand{\dmusyst}{\Delta\mu_{\rm syst-z}}

\newcommand{\alphaTrueSym}{\alpha_{\rm true}}
\newcommand{\betaTrueSym}{\beta_{\rm true}}
\newcommand{\alphaTrueVal}{0.14}
\newcommand{\betaTrueVal}{3.1}

\newcommand{\sigmu}{\sigma_{\mu}}
\newcommand{\sigmubar}{\overline{\sigma_{\mu}}}

\newcommand{\sigz}{\sigma_{z}}
\newcommand{\sigR}{\sigma_{R}}
\newcommand{\sigint}{\sigma_{\rm int}}

\newcommand{\COVsyst}{{\rm COV}_{\rm syst}}
\newcommand{\COVsysti}{{\rm COV}_{{\rm syst},i}}
\newcommand{\COVstat}{{\rm COV}_{\rm stat}}

\newcommand{\wCDM}{$w$CDM}
\newcommand{\wwCDM}{$w_0w_a$CDM}
\newcommand{\ww}{$w_0$-$w_a$}

\newcommand{\NZBIN}{\textcolor{red}{14}}

\newcommand{\AVGwbias}{\langle w$-bias$\rangle}
\newcommand{\AVGwsigbiaszspecsyst}{$0.025$}
\newcommand{\AVGwsigbiaszphotsyst}{$0.023$}
\newcommand{\AVGwsig}{\langle\sigma_w\rangle}
\newcommand{\STDw}{{\rm STD}_w}

\newcommand{\AVGwwbias}{\langle {w_0}$-bias$\rangle}
\newcommand{\AVGwwsig}{\langle\sigma_{w_0}\rangle}
\newcommand{\STDww}{{\rm STD}_{w_0}}

\newcommand{\AVGwabias}{\langle {w_a}$-bias$\rangle}
\newcommand{\AVGwasig}{\langle\sigma_{w_a}\rangle}
\newcommand{\STDwa}{{\rm STD}_{w_a}}

\newcommand{\AVGFoM}{{\langle}{\rm FoM}{\rangle}}

\newcommand{\RatioFoM}{{\cal R}_{{\rm FoM},i}}
\newcommand{\FoM}{{\rm FoM}}
\newcommand{\FoMStat}{{\rm FoM}_{\rm stat}}
\newcommand{\FoMSysti}{{\rm FoM}_{{\rm syst},i}}
\newcommand{\AVGFOMzspecstat}{$136$}
\newcommand{\AVGFOMzspecsyst}{$95$}
\newcommand{\AVGFOMzphotstat}{$237$}
\newcommand{\AVGFOMzphotsyst}{$145$}
\newcommand{\AVGwbiaszphotsyst}{$0.0091$}
\newcommand{\AVGwabiaszphotsyst}{$0.0363$}
\newcommand{\AVGwbiaszspecsyst}{$0.0140$}
\newcommand{\AVGwabiaszspecsyst}{$0.0662$}

\author{Dharmendra Kumar${}^{1}$ } 
\email{12dharmendra13@gmail.com }
\author{Ayan Mitra${}^{2,3}$} 
\email{ayan@illinois.edu}
\author{Shahnawaz A. Adil${}^{4}$ } 
\email{shazadil14@gmail.com}
\author{Anjan A. Sen${}^{5}$ } 
\email{aasen@jmi.ac.in}

\affiliation{
${}^1$Malaviya National Institute of Technology Jaipur-302017, India\\
${}^2$Center for AstroPhysical Surveys, National Center for Supercomputing Applications, University of Illinois Urbana-Champaign, Urbana, IL, 61801, USA\\
${}^3$Department of Astronomy, University of Illinois at Urbana-Champaign, Urbana, IL 61801, USA\\
${}^4$Department of Physics, Jamia Millia Islamia, New Delhi-110025, India \\
${}^5$Centre for Theoretical Physics, Jamia Millia Islamia, New Delhi-110025, India\\
}

\date{\today} 

\begin{abstract}
In recent years, the Lambda Cold Dark Matter ($\Lambda$CDM) model, which has been pivotal in cosmological studies, has faced significant challenges due to emerging observational and theoretical inconsistencies. This paper explores alternative cosmological models to address these discrepancies, using simulated three years photometric Supernovae Ia data from the Legacy Survey of Space and Time (LSST), supplemented with additional Pantheon+, Union, and the recently released Dark Energy Survey 5 Years (DESY5) supernova compilations and Baryon Acoustic Oscillation (BAO) measurements. We assess the constraining power of these datasets on various dynamic dark energy models, including CPL, BA, JBP, SCPL, and GCG. Our analysis demonstrates that the LSST  with its high precision data, can provide tighter constraints on dark energy parameters compared to other datasets. Additionally, the inclusion of BAO measurements significantly improves parameter constraints across all models. 
 Except for Pantheon+, we find that across all the cosmological datasets, and the dark energy models considered in this work, there is a consistent deviation from the $\Lambda$CDM model that exceeds a $2\sigma$ significance level.
Our findings underscore the necessity of exploring dynamic dark energy models, which offer more consistent frameworks with fundamental physics and observational data, potentially resolving tensions within the $\Lambda$CDM paradigm. Furthermore, the use of simulated LSST data highlights the survey's potential in offering significant advantages for exploring alternative cosmologies, suggesting that future LSST observations would play a crucial role.
\end{abstract}

\maketitle
\section{Introduction}
\label{sec:intro}

Over the past two decades, the Lambda Cold Dark Matter ($\Lambda$CDM) cosmological model has been instrumental in explaining a vast array of cosmological observations successfully. Yet, as we've transitioned into an era marked by high-precision cosmology—bolstered by richer, deeper, and higher-resolution data—some cracks in the 
$\Lambda$CDM model have surfaced. Notably, certain discrepancies have emerged that challenge its comprehensiveness, some of which are becoming increasingly significant (\citet{Aluri:2022hzs, Perivolaropoulos:2021jda, Kamionkowski:2022pkx, DiValentino:2021izs,Krishnan:2021dyb,Abdalla:2022yfr,Krishnan:2021jmh,Smith:2022hwi,SolaPeracaula:2022hpd}).

The $\Lambda$CDM model, combining the cosmological constant ($\Lambda$) and cold dark matter (CDM), has long been the cornerstone of cosmological understanding, owing to its strong alignment with a wide range of observational data, such as the Cosmic Microwave Background \cite{planck2020}, Type Ia supernova observations \cite{scolnic2018}, and Baryon Acoustic Oscillations \cite{eisenstein2005}. However, despite its glory, this model faces several significant challenges, both observational and theoretical.

Observationally, the tensions between the values of $H_0$  obtained from CMB observations \cite{Planck2018} and SNIa observations (\cite{pantheon_new, Pantheon, Riess_2022}) pose serious challenges on the viability of the existing $\Lambda$CDM model. The SNIa observations from the SH0ES team \cite{Riess_2022}, provide the most accurate model independent calculation of $H_0$ value using the low-$z$ measurements of local Cepheids
variables in the host SNIa. Which is estimated $\sim 73.30\pm 1.04$ Km/sec/Mpc. This creates a strong $5\sigma$ tension (popularly labelled as the Hubble tension) with the $H_0$ measurement value of $67.36\pm 0.54$ Km/sec/Mpc obtained from the CMB observations \cite{Planck2018}. Besides this strong observational disagreement, there are multiple theoretical disagreements too, on the validity of the $\Lambda$CDM model.

Theoretically, from the standpoint of string theory, the existence of a stable de Sitter (dS) vacuum, which is essential for the $\Lambda$CDM model, is highly problematic. String theory suggests that such vacua might not exist, or are at least extremely difficult to construct \cite{vafa2005, obied2018, ooguri2019}. This fundamental issue raises serious concerns about the feasibility of the $\Lambda$CDM model, which relies on a positive cosmological constant to describe the accelerated expansion of the Universe.

Furthermore, the $\Lambda$CDM model posits a constant dark energy component, $\Lambda$, which is inconsistent with the predictions of quantum field theory (QFT). QFT predicts a vacuum energy density that is many orders of magnitude larger than what is observed, leading to what is known as the cosmological constant problem (\cite{sahni2000, weinberg1989}). This stark discrepancy suggests that the current understanding of $\Lambda$ may be incomplete or fundamentally flawed.

Alternative cosmological models propose dynamic dark energy components, such as quintessence fields, which evolve over time and do not remain constant. These models can naturally arise in an anti-de Sitter (AdS) vacuum, which is more consistent with string theory (\cite{maldacena1999, garg2019}). Unlike de Sitter vacua, there are numerous consistent solutions in an AdS background, making these scenarios theoretically robust (\cite{copeland2006, tsujikawa2013}). This indicates that focusing exclusively on a de Sitter phase might be limiting and that exploring AdS scenarios could provide more viable alternatives.

The empirical success of the $\Lambda$CDM model cannot be overlooked, yet its theoretical underpinnings remain contentious. The reliance on a positive $\Lambda$ leads to a universe that will eventually settle into a de Sitter phase, a scenario fraught with challenges from both string theory and QFT perspectives. Consequently, it is essential to explore alternative cosmological models that involve dynamic dark energy evolving towards an AdS vacuum, as these may offer a more consistent framework with fundamental physics and observational data.

\section{Background}\label{sec:background}
For a homogeneous and isotropic universe, the space-time metric is described by the Friedmann-Lemaître-Robertson-Walker (FLRW) metric:

\[
ds^2 = -dt^2 + a^2(t) \left(dx^2 + dy^2 + dz^2\right),
\]

where, $a(t)$ is the scale factor.\\

Using the FLRW metric, the solution to Einstein's field equations gives the two Friedmann equations:

\begin{eqnarray}
3H^2 = \frac{8 \pi G}{3} \sum_{i} \rho_i, \label{F1}\\
2 \dot{H} + 3 H^2 = - 4 \pi G \sum_{i} p_i,  \label{F2}
\end{eqnarray}

where \(H \equiv \dot{a}/a\) is the Hubble parameter, \(\rho_i\) is the energy density, and \(p_i\) is the pressure for the \(i\)-th fluid. Here, \(i=m\) represents matter density, \(i=r\) represents radiation density, and \(i=de\) represents the total dark energy density in the universe. The continuity equation is thus expressed as:
\begin{equation}
 \dot{\rho_i} + 3 H (\rho_i + p_i) = 0 ,  
\end{equation}

For matter, setting p = 0, we can write $\rho_m = \rho_{m0} a ^{-3}$, Similarly for radiation density $p=\frac{1}{3} \rho$, so $\rho_r = \rho_{r0} a ^{-4}$, and, finally for the dark energy part, we assumed equation of state parameter $w = \frac{p}{\rho}$ and can be written as 
\begin{equation}
\rho_{de} = \rho_{de0} \exp\left(-3\int_{a_{0}}^{a}(1+w)da/a\right),
\end{equation}

And we can also write  Hubble parameter as,

\begin{equation}
H^{2}(z) = H_{0}^2 \left(  \Omega_{m} (1+z)^{3} + \Omega_{r} (1+z)^{4} +  (1-\Omega_{m} - \Omega_{r}) \exp\left[3 \int \frac{1+w}{1+z} dz \right]   \right).
\label{eq:hubble}
\end{equation}

where $\Omega_{m} = \frac{8\pi G}{3H^2_0} \rho_{m0}$, $\Omega_{r} = \frac{8\pi G}{3H^2_0} \rho_{r0}$ and $\Omega_{de} = \frac{8\pi G}{3H^2_0} \rho_{de0}$, are the present value of matter density, radiation density, dark energy density respectively with the flatness condition $\Omega_{m} +\Omega_{r}+\Omega_{de} = 1 $ and $H_0$ is Hubble constant, present value of Hubble parameter H(z).


\subsection{Dark Energy Models}
\label{sec:demodels}
Now to solve equation \ref{eq:hubble}, we need to consider some form of equation of state for dark energy ($w$). Here, we consider five widely used parameterizations of $w$ which are : CPL~\citet{Chevallier:2000qy,Linder:2002et}, BA~\citet{2008PhLB..666..415B}, JBP~\citet{PhysRevD.67.063504}, SCPL~\citet{2016PhRvD..93j3503P}, and GCG~\citet{Thakur:2012rp}. The form of $w$ for these parameterizations are as follows:

\begin{eqnarray}
w(z) &=& w_{0} + w_{a}\frac{z}{1+z} \hspace{3mm} (CPL) \hspace{1mm}\nonumber\\
w(z) &=& w_{0} + w_{a} \frac{z(1+z)}{1+z^2} \hspace{3mm} (BA) \hspace{1mm}\nonumber\\
w(z) &=& w_{0} +w_{a}\frac{z}{(1+z)^2} \hspace{3mm} (JBP) \hspace{1mm}\\
w(z) &=& w_{0} + w_{a} \left(\frac{z}{1+z}\right)^7 \hspace{3mm} (SCPL) \hspace{1mm}\nonumber\\
w(z) &=& -\frac{w_{0}}{w_{0} + (1-w_{0}) (1+z)^{3 (1+w_{a})}} \hspace{3mm} (GCG) \hspace{1mm} \nonumber
\end{eqnarray}

\subsection{Supernova Ia}
Type Ia supernovae (SN Ia) are believed to be the result of the thermonuclear disruption of
a carbon-oxygen white dwarfs which reaches the Chandrasekhar-mass limit of stability ($M_{Ch} \sim 1.4$M solar mass) by accreting matter from a companion ~\citet{1960ApJ...132..565H}. SNIa light curves show remarkable homogeneity after correcting for stretch and color parameters. Observed flux from SNIa can be used to compute the luminosity distance $D_L$. For an FLRW Universe, with a dark energy component EoS $w(z)$ the $D_L$ is given as,  
\begin{equation}
D_L =  (1+z) \frac{c}{H_0} \int_{0}^{z} \frac{dz'}{E(z')}
\end{equation}
where distance $D_L$ is in megaparsecs (Mpc) and $E(z')$ is normalised Hubble parameter ($H(z)/H_0$) given as,
\begin{equation}
    E(z) = \sqrt{\Omega_m (1+z)^3 + (1 - \Omega_m) \exp\left(3 \int_{0}^{z} \frac{1 + w(z')}{1 + z'} dz'\right)}
\end{equation}
and the distance modulus $\mu$ can be related to the $D_L$ (for a flat Universe with a constant EoS dark energy $w=-1$) as 
\begin{equation}
\mu = 5\log (D_L /10\text{pc}),
\end{equation}
 therefore SNIa provides a direct and robust method to probe dark energy. Over the last two decades, numerous studies have utilized Type Ia Supernovae (SNIa) as the primary probe for investigating dark energy ~\citet{panstarrs,Betoule2014,s1,Dhawan:2021ztf}. 
 Through successive SNIa programs, our precision in understanding dark energy has significantly improved, with efforts still ongoing ~\citet{Kessler2010, Linder2019,Mitra2021}.


\section{LSST: Overview}\label{sec:lsst}
The Legacy Survey of Space and Time (LSST), conducted by the Vera C. Rubin Observatory collaboration, represents a paradigm shift in astrophysical surveys with its unparalleled scope and technical sophistication. Scheduled to begin in 2023, the LSST will deploy an advanced observational apparatus, featuring an $8.4$ m primary mirror with a $6.7$ m effective aperture and a state-of-the-art $3200$-megapixel camera, yielding a wide field of view of $9.6$ square degrees. The LSST is designed to survey approximately 18,000 square degrees of the southern sky over a decade, utilizing six optical passband filters to facilitate deep, wide, and fast observations. The survey aims to amass over 32 trillion observations of 20 billion galaxies and a similar number of stars, achieving a depth of $24^{th}$ magnitude in its six filter bands, spanning wavelengths from ultraviolet to near-infrared ~\citet{Cahn2009, ivezic}. These efforts are projected to catalog millions of supernovae, among other transient phenomena, offering an unprecedented dataset for probing the dark Universe. The Dark Energy Science Collaboration (DESC) \footnote{\url{https://lsstdesc.org/}}, comprising nearly 1,000 members, intends to harness this vast trove of data to extract high-precision measurements of fundamental cosmological parameters, leveraging prior data challenges to refine analysis pipelines in anticipation of the survey's extensive data output ~\citet{dc2, Sanchez}. 

\section{Datasets}\label{sec:datasets}

\subsection{Simulated LSST SNIa Data}\label{sec:sim}
In this analysis, we used simulated LSST SNIa data as the primary data to compare with our existing results and the dark energy constraining powers with different dark energy models listed above. The details of the simulation is based on ~\citet{Mitra:2022ykq}.

The SN data is generated using the LSST Dark Energy Science Collaboration (DESC) 
  time domain (TD) pipeline and \SNANA\ code ~\citet{snana}, consisting of four main stages illustrated in Fig. 3 
of ~\citet{Mitra:2022ykq}. This includes SN brightness standardisation via a Light Curve (LC) fit stage, simulations for bias correction, and a BBC stage for Hubble diagram production before the last stage, cosmology fitting.    We used a redshift binned Hubble diagram and the associated covariance matrix produced from the BBC stage to perform cosmological fitting. 

A brief explanation of how the  LSST SNIa mock data are composed, is given below (more details can be found in sec.~4 of ~\citet{snana_manual}). An input cosmology corresponding to a model is provided, for standard cosmology, the default is $\Omega_m=0.3150, \, \Omega_\Lambda=0.6850, \, w_0=-1, \, w_a=0$. The curvature is computed internally as $\Omega_k=1-\Omega_m-\Omega_\Lambda$.  It is based on cosmological parameters from Planck 2018 ~\citet{Planck2018}. In addition, the parameter $H_0$ is set to $70.0$ km/s/Mpc, this value is tied to SALT2 training. SALT2 lightcurve model generates observer frame magnitudes. The noise in the simulation is computed as follows : 
\small
\begin{equation}
    \sigma_{\text{SIM}}^2 = \left[ F + (A\cdot b) + (F\cdot \sigma_{\text{ZPT}})^2 + \sigma_0\cdot 10^{0.4\cdot ZPT_{\text{pe}}} + \sigma_{\text{host}}^2 \right]S^2_{\text{SNR}}.
\end{equation}
\normalsize

where $F$ is the simulated flux in photoelectron (p.e.), $A$ is the noise equivalent area given by, $A = \left[2\pi\int \text{PSF}^2 (r,\theta)r \mathrm{d} r\right]^{-1}$ where PSF stands for the Point Spread Function, $b$ is the background per unit area (includes sky + CCD readouts + dark current), $S_{\text{SNR}}$ is en empirically determined scale that depends on the signal-to-noise ratio, the three $\sigma$ terms correspond to zero point uncertainty, flux calibration uncertainty and underlying host galaxy uncertainty. These terms can be determined empirically from fits that match simulated uncertainties to those from the survey designs.

\begin{figure*}
\includegraphics[scale=0.6]{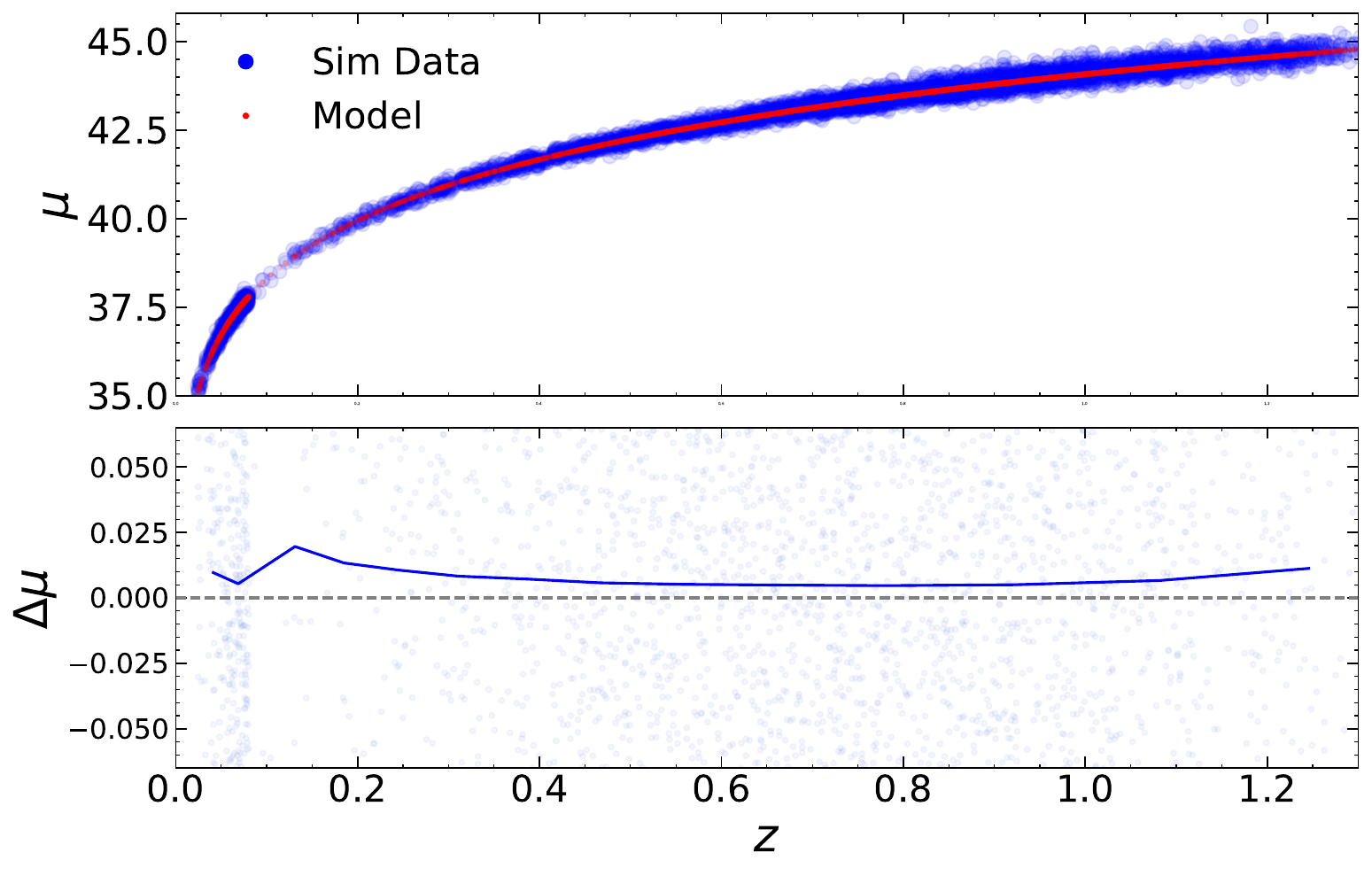}
    \caption{Plot illustrating the Hubble diagram for wCDM  cosmology model with simulated LSST data. The bottom panel shows the Hubble residual plot with respect to true cosmology. }
    \label{fig:hd_LSST}
\end{figure*}

The SNIa dataset used is composed of spectroscopically ($\zspec$) and photometrically ($\zphot$) identified SNIa candidates combined together. Based on \acro\ ~\citet{plasticc_announce} ``$\zspec$'' sample is composed of two sets of events whose spectroscopic redshifts have an accuracy of $\sigma_{z} \sim 10^{-5}$. The first subset is made up of \specy\ confirmed events whose redshift is predicted to be accurate by the 4MOST spectrograph ~\citet{4MOST2}, which is being built by the European Southern Observatory and is expected to become operational in $2023$. 4MOST is situated at a latitude similar to that of the Rubin Observatory in Chile. The second subset is composed of photometrically identified events with an accurate host galaxy redshift determined by 4MOST. The second subset has about $\sim60$\% more events than the first subset.
For the photometric sample, host galaxy photo-$z$  was used as a prior (adapted from ~\citet{Kessler2010}).  The photometric redshift and rms uncertainty was based on ~\citet{Graham2018_photoz}. The whole simulation was re-done based on the \acro \  DDF data\footnote{The LSST has different observing strategies, the deep field and the wide field, called as the DDF (Deep drilling field) and the WFD (Wide Fast Deep) respectively.}. Additional low redshift spectroscopic data was used from the DC2 analysis (simulated with WFD cadence). ~\citet{Mitra:2022ykq} simulated only SNIa and no contamination (eg. core collapse, peculiar SNe etc) was considered. The covariance matrix used corresponds to the stat+syst where syst is all individual systematics combined. Seven systematics in total were considered, which are detailed in Table 3 of ~~\citet{Mitra:2022ykq}. The Hubble diagram is unbinned and contains a total of  $5784$ SNIa candidates, as shown in Fig.~\ref{fig:hd_LSST}, composed of both spectroscopic and photometric candidates. 

\subsection{Pantheon+ compilations}
 We use the SNe Ia distance moduli measurements from the Pantheon+ sample ~\citet{Brout_2022}, which consists of 1701 light curves of 1550 distinct SNe Ia ranging in the redshift interval $z \in [0.001, 2.26]$. We refer to this dataset as \texttt{Pantheon-Plus}. We also consider the SH0ES Cepheid host distance anchors, which gives constraints on both $M_B$ and $H_0$. When utilizing SH0ES Cepheid host distances, the SNe Ia distance residuals are modified following the relationship eq.(14) of   ~\citet{Brout_2022}. 
\subsection{DESY5 compilations}
We use Type Ia supernovae (SN Ia) discovered and measured over the full five years of the Dark Energy Survey (DES) Supernova Program. This includes 1635 DES SNe in the redshift range $0.10 < z < 1.13$ that meets the quality selection criteria and can be used to constrain cosmological parameters. To calibrate these 1635 SNe, we have used a high-quality external low-redshift sample consisting of 194 SNe Ia spanning $0.025 < z < 0.10$. Consequently, we utilize 1829 distance modulus observations of SN Ia from the DESY5 compilation~\citet{DESY5}.
\subsection{UNION 3.0 compilations}
We use updated “Union” compilation of 2087 cosmologically useful SNe Ia
from 24 datasets (“Union3”). In our analysis we have used 22 binned modulas distance observations of Union3 compilations~\citet{UNION3}. 

\subsection{BAO 2021 compilations}
From the latest compilation of  Baryon Acoustic Oscillation (BAO) distance and expansion rate measurements from the SDSS collaboration, we use 14 BAO measurements,which is, the isotropic BAO measurements of $D_V(z)/r_d$ (where $D_V(z)$ and $r_d$ are the spherically averaged volume distance, and sound horizon at baryon drag, respectively) and anisotropic BAO measurements of $D_M(z)/r_d$ and $D_H(z)/r_d$ (where $D_M(z)$ and $D_H(z)=c/H(z)$ are the comoving angular diameter distance and  the Hubble distance, respectively), as compiled in Table 3 of ~\citet{Alam_2021}. 

In the likelihood analysis of the different parameters we have used uniform priors. We have used Markov Chain Monte Carlo (MCMC) method to make the parameter estimation. We used MCMC sampler, EMCEE,~\citet{Goodmanetal2010,Foremanetal2013} in {\tt Python}. We have studied these MCMC chains using the program GetDist~\citep{Lewis}.


\section{Results \& Discussions}\label{sec:results} 
In this section, we present the outcomes of our analysis. We explored six distinct dark energy models detailed in subsection \ref{sec:demodels}, aiming to compare their impacts using various observational datasets including Pantheon-Plus, LSST, Union3, and DESY5, with and without the BAO 2021 compilations. Throughout our analysis, we fixed the values of $\Omega_m$ and $H_0$ at $0.3$ and $70$ respectively, and examined all dark energy models across the $w_0-w_a$ plane.

When considering LSST data without BAO constraints, for the CPL model, we find $w_0$ and $w_a$ to be $-0.929\pm 0.025$ and $-0.20\pm 0.16$ respectively. The error bars were $2.7\%$ for $w_0$ and $80\%$ for $w_a$. Similarly, for the BA model, the values were $-0.935\pm 0.020$ and $-0.24\pm 0.20$ with error bars of $2.1\%$ for $w_0$ and $83\%$ for $w_a$. Furthermore, for the JBP model, constraints were $-0.920\pm 0.031$ and $-0.34\pm 0.27$, with error bars of $3.4\%$ and $79.4\%$ respectively. For the SCPL model, we obtained constraints of $-0.959\pm 0.005$ for $w_0$ and $-0.10^{+1.8}_{-3.7}$ for $w_a$, with error percentages of $0.52\%$ and inconclusive constraint for $w_a$ as depicted in Figure \ref{fig:SCPL ALL data with BAO}. Finally, for the GCG-Thawing model, we find $w_0$ and $w_a$ to be $0.932\pm 0.014$ and $-2.21^{+0.28}_{-0.77}$ respectively, with error percentages of $15\%$ and $37.1\%$ respectively. Additionally, for the GCG-Tracker model, the values are $0.966^{+0.006}_{-0.007}$ and $-0.637^{+0.084}_{-0.35}$ respectively, with error percentages of $0.01\%$ and $56.5\%$ respectively. Incorporating the BAO 2021 compilations did not significantly change the parameter values for this dataset. This could be attributed to tighter constraints from LSST only which is dominating the constraints on these parameters.

For Pantheon-Plus compilations, we find the constraints on $w_0$ as $10.9\%$ , $8.9\%$, $13.04\%$, $4.55\%$, $9.44\%$ and $5\%$ for CPL, BA, SCPL, GCG-Thawing and GCG-Tracker models, respectively. The constraint on $w_a$ parameter in all of these models are quite large. The addition of BAO 2021 compilations significantly improves the constraints on these parameters for all the model as shown in Table \ref{tab:all model}. Similarly, for DESY5 data, the constraints on $w_0$ are $7.27\%$, $5.8\%$, $8.7\%$, $1.5\%$ and $3.25\%$ and $1.82\%$ for CPL, BA, JBP, SCPL, GCG-Thawing and GCG-tracker models respectively. Lastly, We analysed all these models with Union3 data. Here we find  constraints on $w_0$ as $21.73\%$, $16.44\%$, $38.14\%$, $7.53\%$, $13.4\%$ and $9.78\%$ respectively for CPL, BA, JBP, SCPL, GCG-Thawing and GCG-Tracker models. It is important to note that addition of BAO 2021 compilations dataset significantly improves the constraints for all these mentioned models. For this dataset as well, the constraints on $w_a$ are very large for all the models analysed. The constraints on the parameters are presented in Table \ref{tab:all model} for reference. Also, We have presented the $2\sigma$ posteriors for all the relevant models using different dataset in Figures \ref{fig:CPL_all_combined}, \ref{fig:BA_all_combined},\ref{fig:JBP_ALL_combined}, \ref{fig:SCPL ALL data with BAO},\ref{fig:GCG-thawing ALL data},\ref{fig:GCG-track ALL data with BAO}. The red solid lines in these figures represent the constraints from LSST data. We find all the dark energy models have a very tight constraint from LSST data as compared to rest of the data. This could be attributed to very large number of data points (which is $5784$) from LSST measurements.

\subsection{Comparisons with $\Lambda$CDM model} \label{subsec:comparisons}
\begin{figure*}[htb!]
	\centering
	\begin{minipage}{0.45\textwidth}
		\centering
		\includegraphics[width=\linewidth]{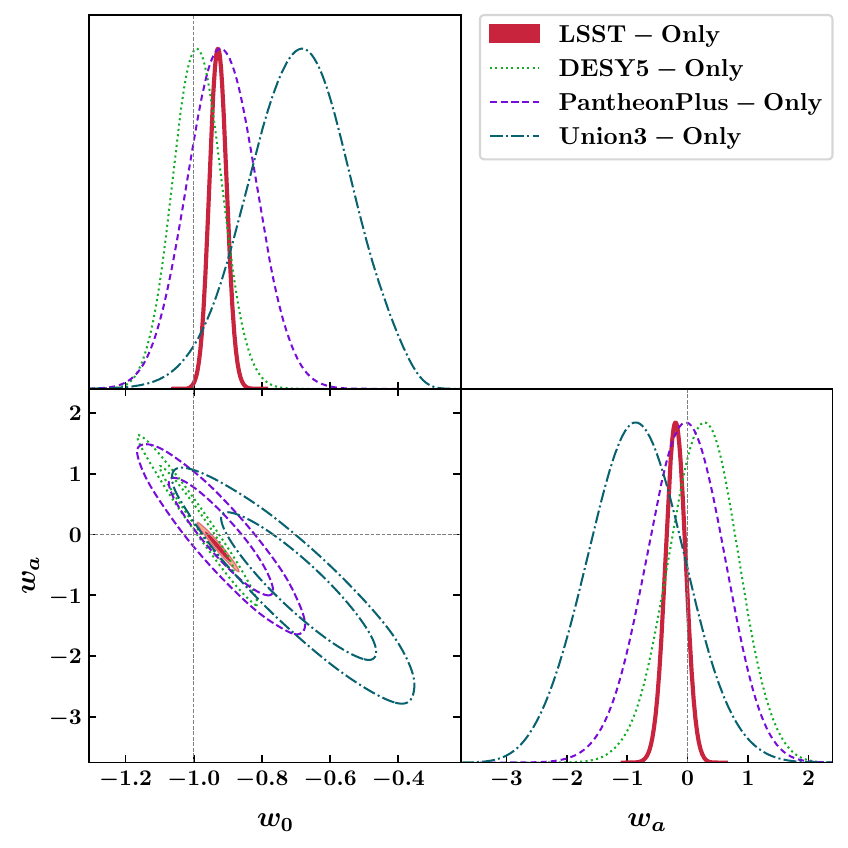}
	\end{minipage}
	\hfill
	\begin{minipage}{0.45\textwidth}
		\centering
		\includegraphics[width=\linewidth]{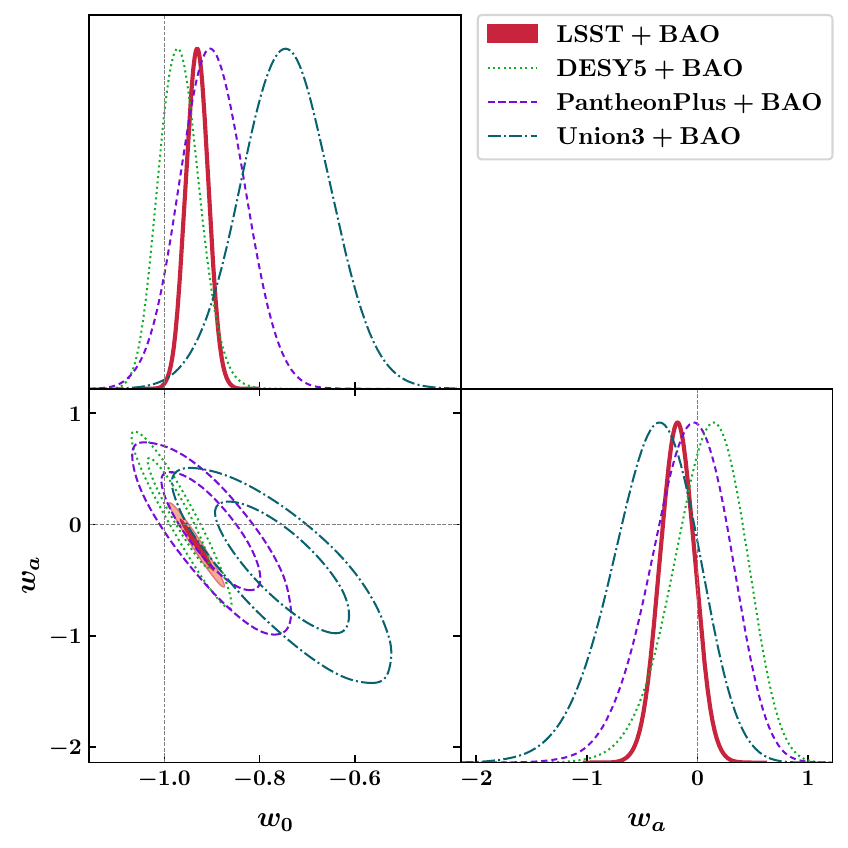}
	\end{minipage}
	\caption{Left panel: CPL with LSST, DESY5,PantheonPlus and Union3 Data. Right panel:CPL with LSST+BAO, DESY5+BAO, PantheonPlus+BAO and Union3+BAO data. Where black dashed line represents the $\Lambda$CDM point (fiducial cosmology, $w_0,w_a=-1,0$). All dark energy contours shown (here and after) are $2\sigma$ constraint ellipses.}
         \label{fig:CPL_all_combined}
\end{figure*}

\begin{figure*}[htb!]
	\centering
	\begin{minipage}{0.45\textwidth}
		\centering
		\includegraphics[width=\linewidth]{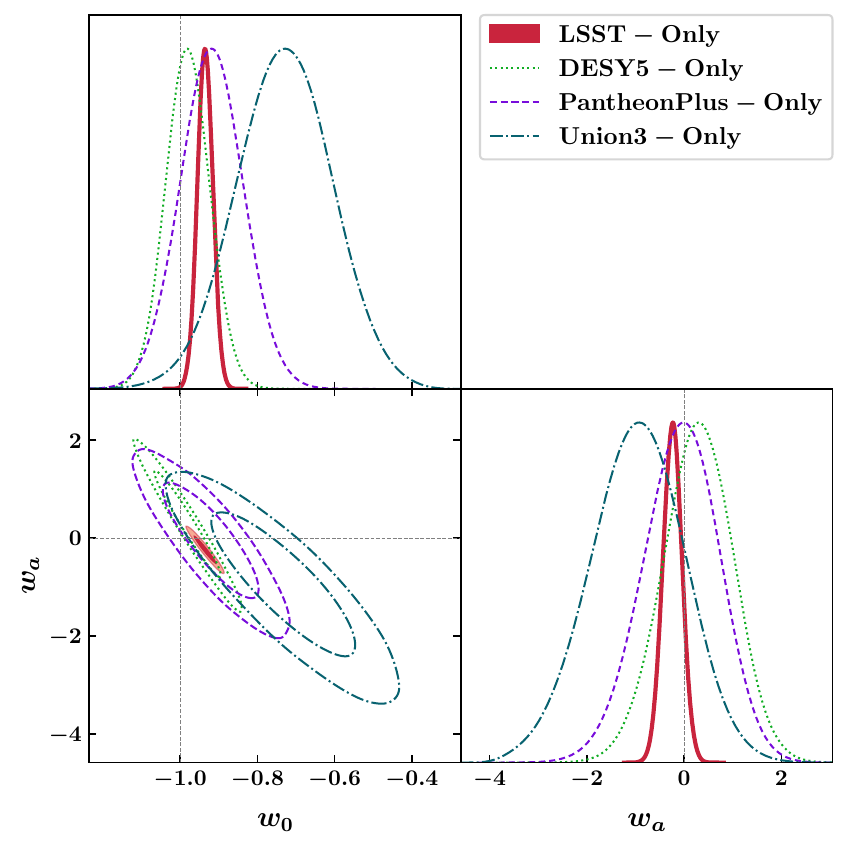}
	\end{minipage}
	\hfill
	\begin{minipage}{0.45\textwidth}
		\centering
		\includegraphics[width=\linewidth]{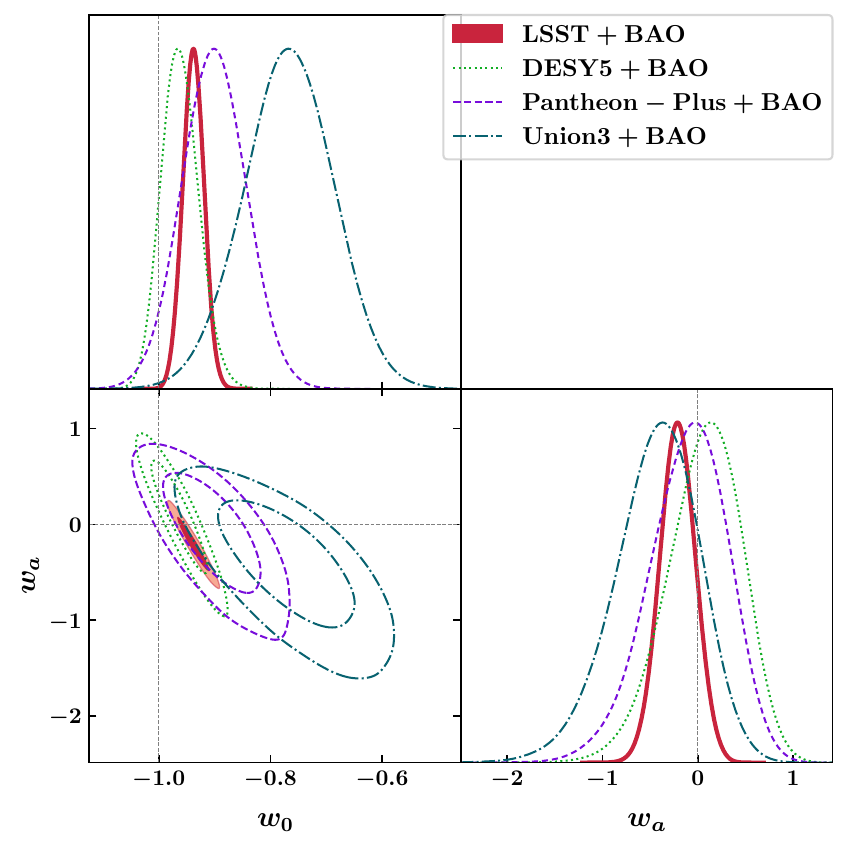}
	\end{minipage}
	\caption{Left panel: BA with LSST, DESY5, PantheonPlus and Union3 Data. Right panel:BA with LSST+BAO, DESY5+BAO, PantheonPlus+BAO and Union3+BAO data. Where black dashed line represents the $\Lambda$CDM point.}
        \label{fig:BA_all_combined}
\end{figure*}

\begin{figure*}[htb!]
	\centering
	\begin{minipage}{0.45\textwidth}
		\centering
		\includegraphics[width=\linewidth]{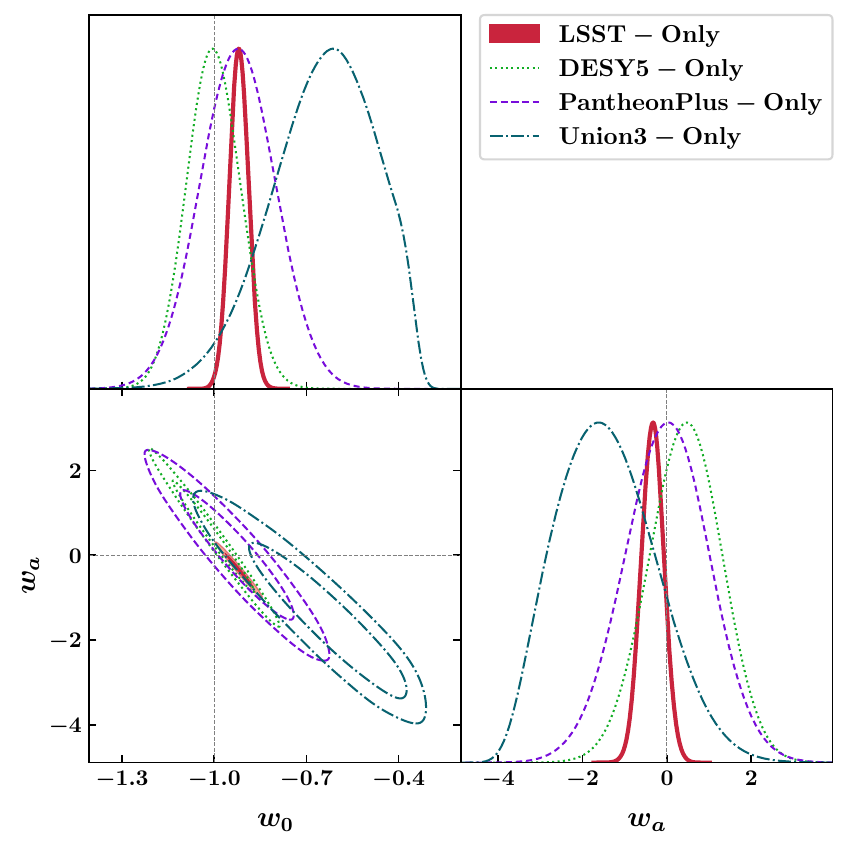}
	\end{minipage}
	\hfill
	\begin{minipage}{0.45\textwidth}
		\centering
		\includegraphics[width=\linewidth]{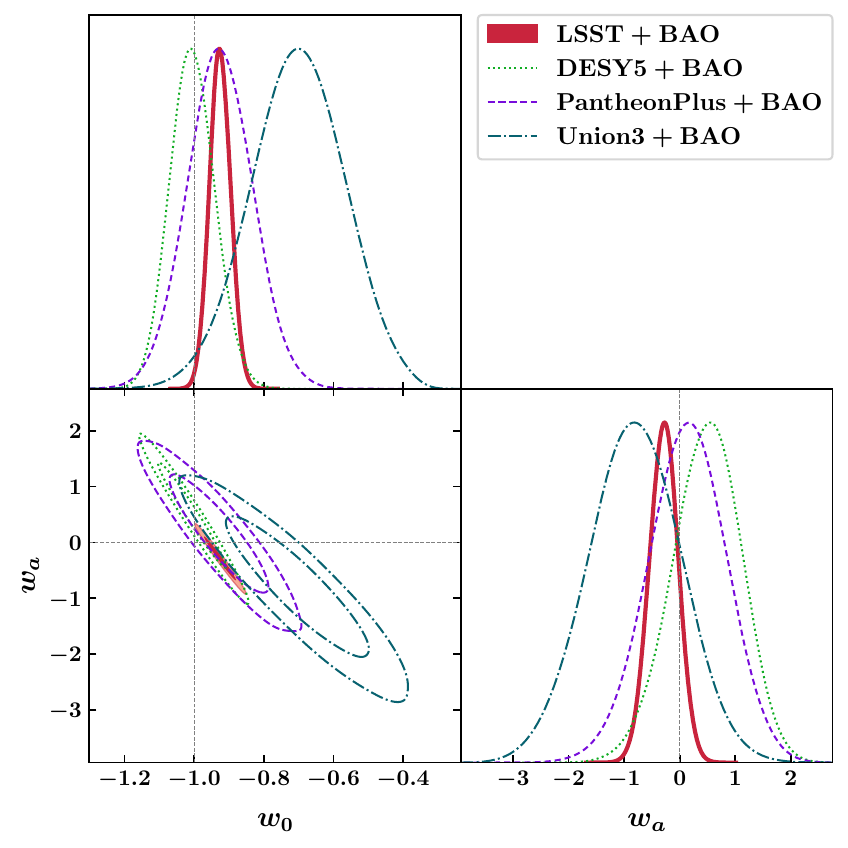}

	\end{minipage}
	\caption{Left panel: JBP with LSST, DESY5, PantheonPlus and Union3 Data. Right panel:JBP with LSST+BAO, DESY5+BAO, PantheonPlus+BAO and Union3+BAO data. Where black dashed line represents the $\Lambda$CDM point.}
        \label{fig:JBP_ALL_combined}
\end{figure*}

\begin{figure*}[htb!]
	\centering
	\begin{minipage}{0.45\textwidth}
		\centering
		\includegraphics[width=\linewidth]{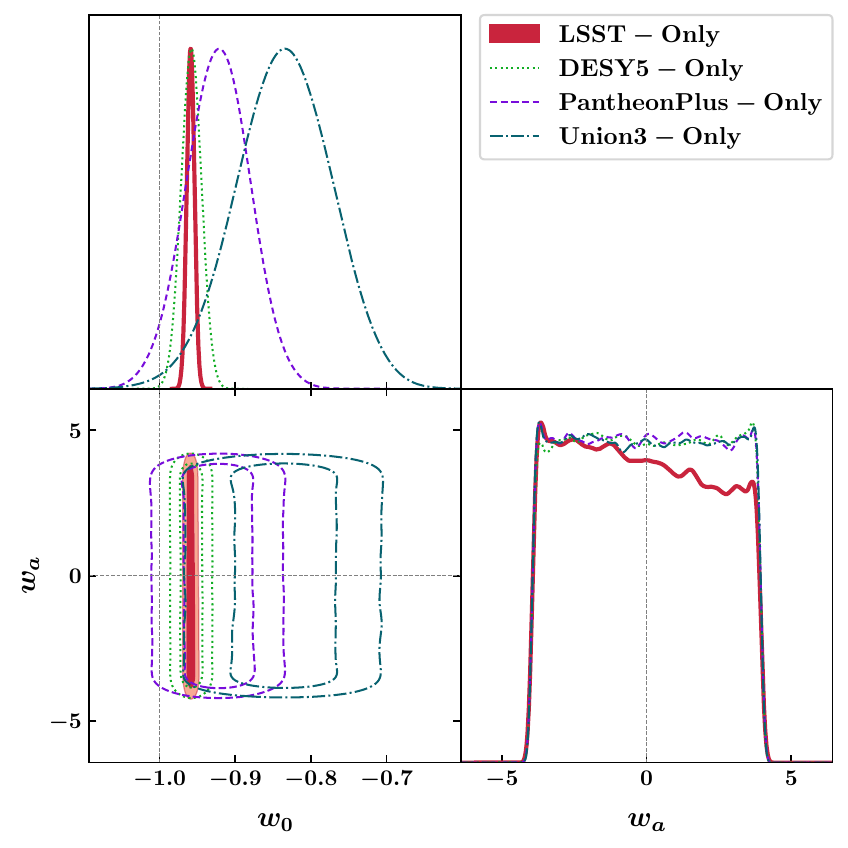}
	\end{minipage}
	\hfill
	\begin{minipage}{0.45\textwidth}
		\centering
		\includegraphics[width=\linewidth]{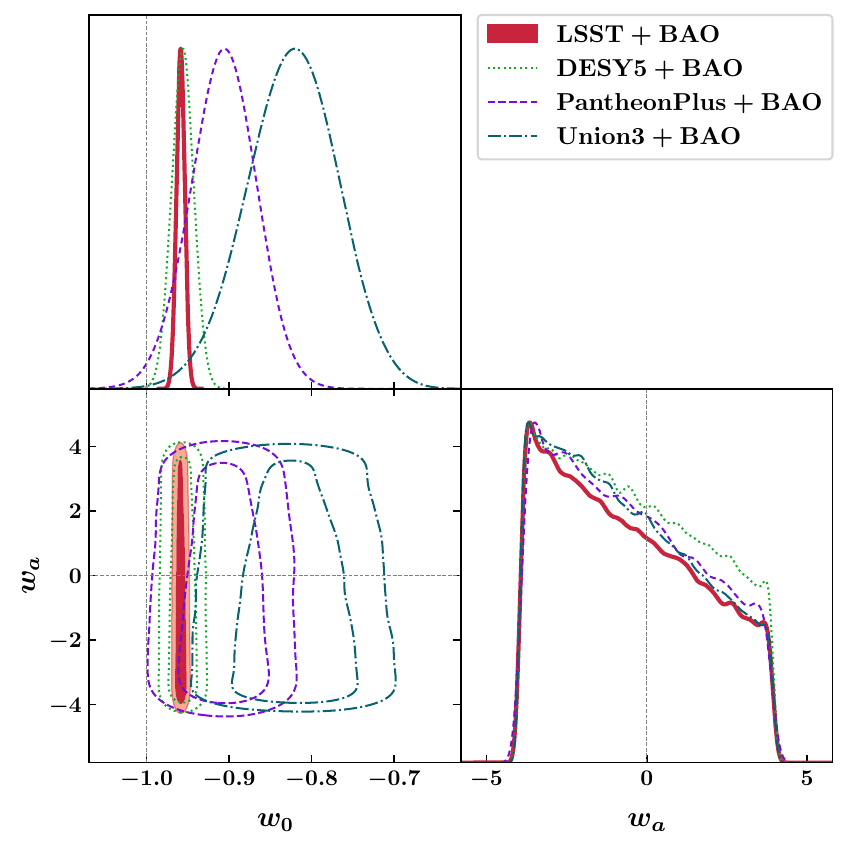}
	\end{minipage}
	\caption{Left panel: SCPL with LSST, DESY5, PantheonPlus and Union3 Data. Right panel: SCPL with LSST+BAO, DESY5+BAO, PantheonPlus+BAO and Union3+BAO data. Where black dashed line represents the $\Lambda$CDM point.}
 \label{fig:SCPL ALL data with BAO}
\end{figure*}

\begin{figure*}[htb!]
	\centering
	\begin{minipage}{0.45\textwidth}
		\centering
		\includegraphics[width=\linewidth]{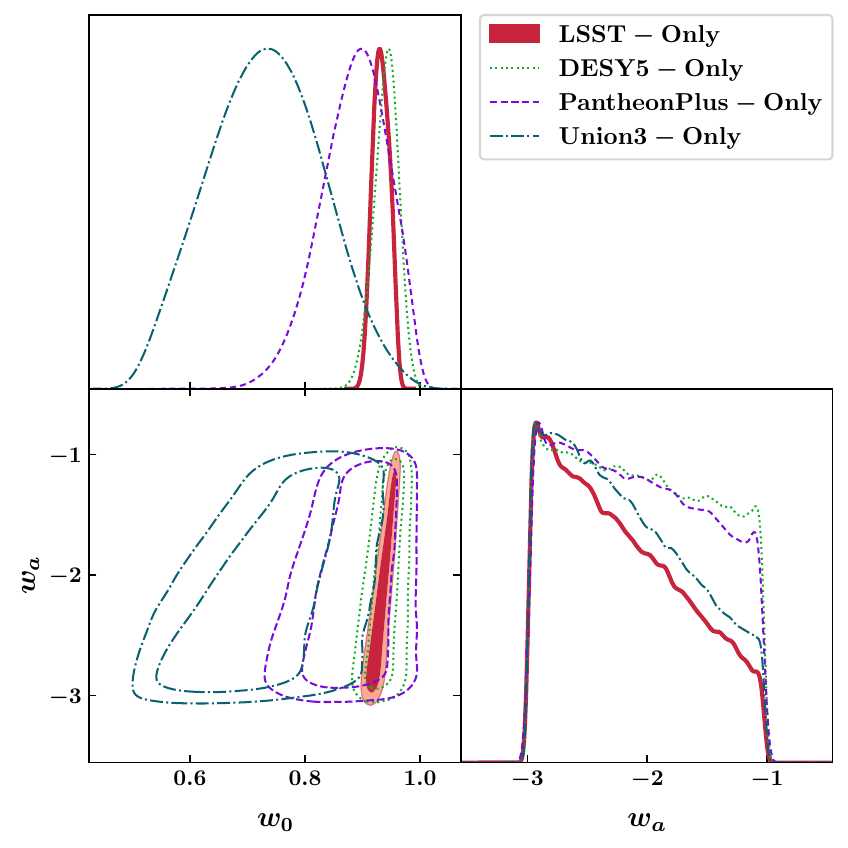}
	\end{minipage}
	\hfill
	\begin{minipage}{0.45\textwidth}
		\centering
		\includegraphics[width=\linewidth]{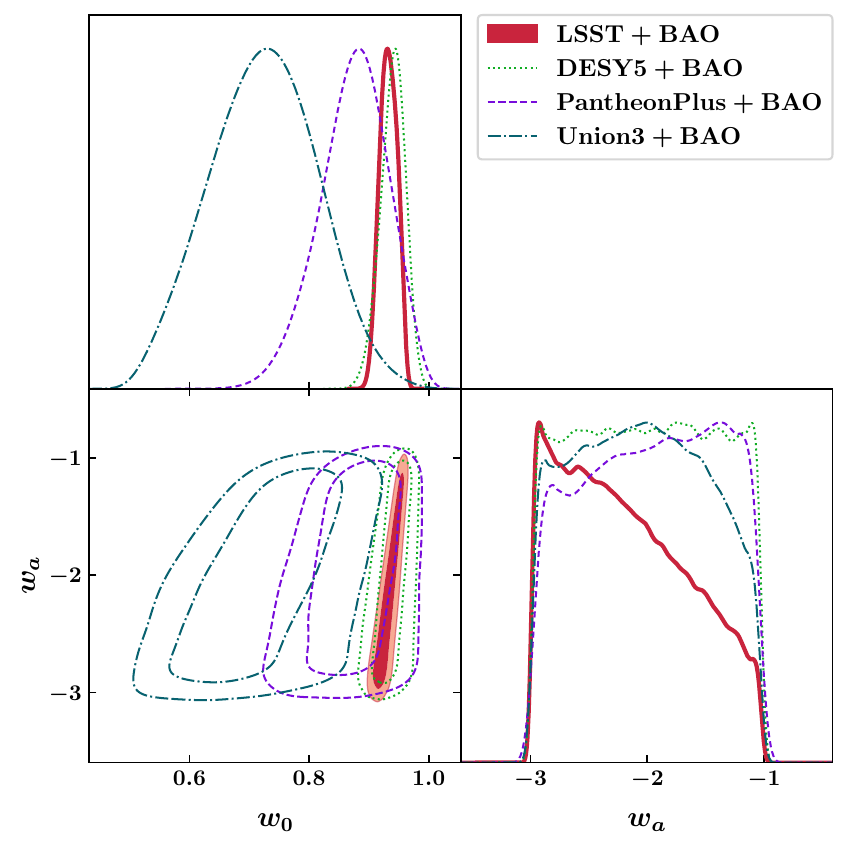}

	\end{minipage}
	\caption{Left panel: GCG-Thawer with LSST, DESY5, PantheonPlus and Union3 Data. Right panel:GCG-Thawer with LSST+BAO, DESY5+BAO, PantheonPlus+BAO and Union3+BAO data. Where black dashed line represents the $\Lambda$CDM point.}
	\label{fig:GCG-thawing ALL data}
\end{figure*}

\begin{figure*}[htb!]
	\centering
	\begin{minipage}{0.45\textwidth}
		\centering
		\includegraphics[width=\linewidth]{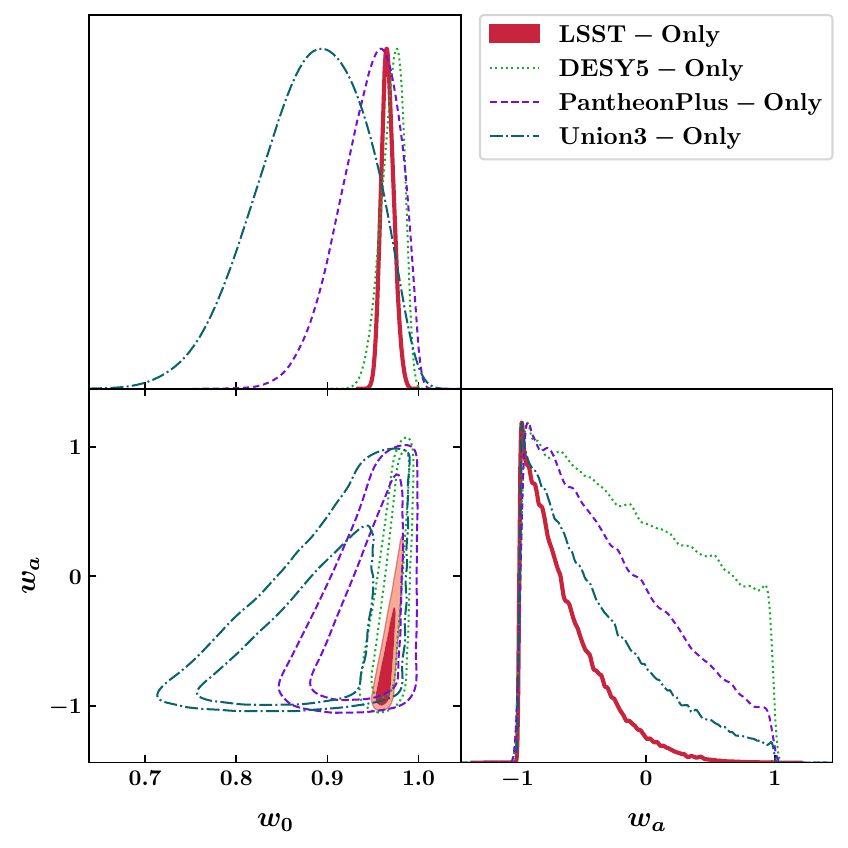}
	\end{minipage}
	\hfill
	\begin{minipage}{0.45\textwidth}
		\centering
		\includegraphics[width=\linewidth]{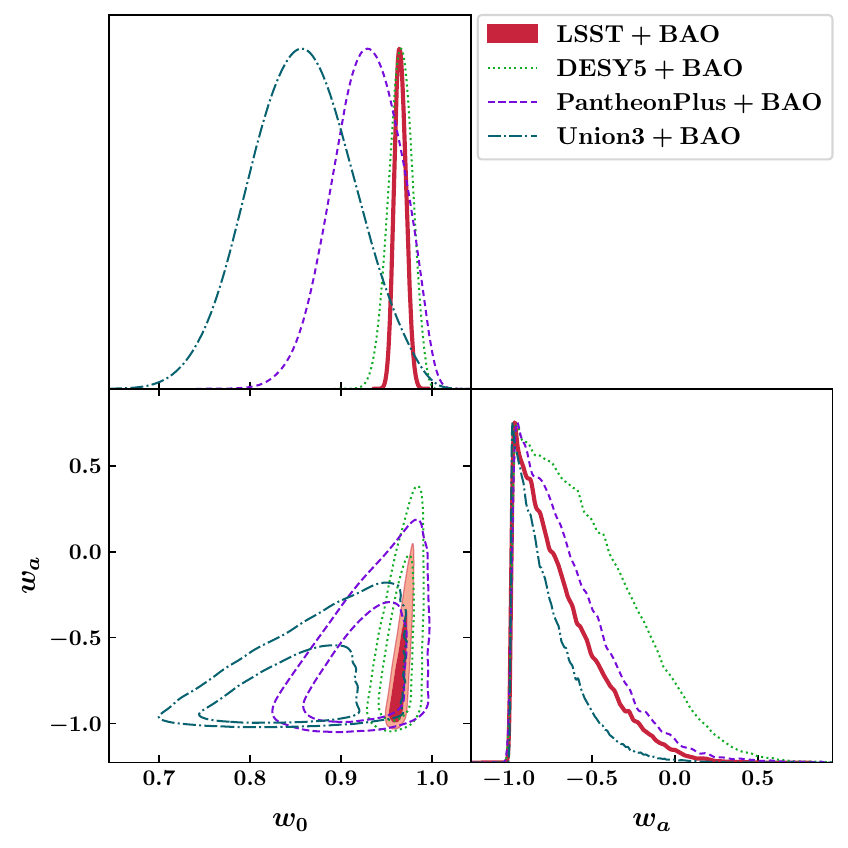}

	\end{minipage}
	\caption{Left panel: GCG-Tracker with LSST, DESY5,PantheonPlus and Union3 Data. Right panel:GCG-Tracker with LSST+BAO, DESY5+BAO, PantheonPlus+BAO and Union3+BAO data. Where black dashed line represents the $\Lambda$CDM point.}
	\label{fig:GCG-track ALL data with BAO}
\end{figure*}

It is interesting to note that in all models and observational data sets, we find the equation of state to be non-phantom, except for the JBP model, particularly with the DESY5 data, where we observe a phantom equation of state. In Figure \ref{fig:CPL_all_combined}, the CPL model shows deviations from the standard $\Lambda$CDM model exceeding $2\sigma$ for the LSST, DESY5, and Union3 data compilations. For the CPL model with Pantheon-Plus data, the deviation is between $1\sigma$ and $2\sigma$. Additionally, in Figure \ref{fig:CPL_all_combined}, we see that adding BAO 2021 data does not reduce the discrepancies with $\Lambda$CDM for the CPL model. In the case of the BA model, as shown in Figure \ref{fig:BA_all_combined}, the deviations persist similarly to the CPL model. The LSST, DESY5, and Union3 compilations show deviations exceeding $2\sigma$, while the Pantheon-Plus data shows a $1\sigma$ deviation. Adding the BAO 2021 compilations to this model increases the deviations, highlighting the tension with the $\Lambda$CDM model. We observe a similar trend for the rest of the models, as illustrated in Figures \ref{fig:JBP_ALL_combined}, \ref{fig:SCPL ALL data with BAO}, \ref{fig:GCG-thawing ALL data}, and \ref{fig:GCG-track ALL data with BAO}.

\begin{table}[ht!]
\centering
\caption{Results for five different cosmological models within 1$\sigma$ limit. }
\begin{tabular}{c c c | c c}

\Xhline{3\arrayrulewidth}
&\textbf{LSST}& & \hspace{0.5in}   \textbf{LSST+BAO}& \\
\Xhline{3\arrayrulewidth}
\textbf{Models} & \textbf{$w_0$} & \textbf{$w_a$}& \textbf{$w_0$} & \textbf{$w_a$}\\ 
\Xhline{3\arrayrulewidth}
CPL &$-0.929\pm 0.025$ & $-0.20\pm 0.16$ &$ -0.931\pm 0.024$ & $-0.18\pm 0.16$\\
BA & $-0.935\pm 0.020 $& $-0.24\pm 0.20$ & $-0.938\pm 0.019$ & $-0.21\pm 0.19$\\
JBP &$-0.920\pm 0.031$ &$ -0.34\pm 0.27 $& $-0.925\pm 0.030 $& $-0.29\pm 0.26$\\
SCPL &$-0.959\pm 0.005$ & $-0.10^{+1.8}_{-3.7}$ & $-0.958\pm 0.005$ & $-0.6^{+1.3}_{-3.3}$\\
GCG-thawer & $0.932\pm 0.014$ & $-2.21^{+0.28}_{-0.77}$ & $0.933\pm 0.014$ & $-2.18^{+0.29}_{-0.80}$\\
GCG-tracker &$0.966^{+0.006}_{-0.007}$ & $-0.637^{+0.084}_{-0.35}$ & $0.965^{+0.006}_{-0.006}$ & $-0.692^{+0.079}_{-0.30}$\\

\Xhline{3\arrayrulewidth}
&\textbf{PantheonPlus}& & \hspace{0.5in}   \textbf{PantheonPlus+BAO}& \\
\Xhline{3\arrayrulewidth}
CPL &$-0.92\pm 0.10 $& $-0.05\pm 0.64 $& $-0.899\pm 0.68 $& $-0.08^{+0.39}_{-0.32}$\\
BA & $-0.919\pm 0.082$ &$ -0.07^{+0.82}_{-0.74}$ & $-0.901\pm 0.058$ & $-0.10^{+0.45}_{-0.37}$ \\
JBP &$-0.92\pm 0.12 $& $-0.0\pm 1.0$ &$ -0.926\pm 0.090$ & $-0.13\pm 0.71$\\
SCPL &$-0.923\pm 0.042 $& $-0.0 \pm 2.3$ & $-0.907\pm 0.040$ & $-0.6^{+1.3}_{-3.3}$\\
GCG-thawer &$0.885^{+0.067}_{-0.050}$ & $-2.08^{+0.36}_{-0.89}$ & $0.874^{+0.064}_{0.050}$ & $-1.97^{+0.85}_{-0.41}$\\
GCG-tracker &$0.943^{+0.042}_{-0.022}$ & $-0.28^{+0.23}_{-0.69}$ & $0.927^{+0.041}_{-0.034}$ & $-0.648^{+0.094}_{-0.34}$\\
\Xhline{3\arrayrulewidth} 
&\textbf{DESY5 }& & \hspace{0.5in}   \textbf{DESY5+BAO}& \\
\Xhline{3\arrayrulewidth}
CPL &$-0.990\pm 0.072 $& $-0.26\pm 0.57$ & $-0.968^{+0.039}_{0.046}$ & $-0.09^{+0.36}_{-0.29}$\\
BA & $-0.980\pm 0.057$ &$ 0.28\pm 0.72$ & $-0.962^{+0.031}_{-0.036}$ & $0.06^{+0.43}_{-0.34}$ \\
JBP &$-1.002\pm 0.087$ & $0.43\pm 0.87$ & $-1.005\pm 0.064$ & $0.48^{+0.65}_{-0.58}$ \\
SCPL &$-0.958\pm 0.014$ &$ -0.0 \pm 2.3 $& $-0.956\pm 0.013$ & $-0.4^{+1.4}_{-3.5}$\\
GCG-thawer &$0.941^{+0.024}_{-0.019}$ &$ -2.05^{+0.57}_{-0.92}$ & $0.940^{+0.023}_{-0.018}$  & $-2.00 \pm 57$ \\
GCG-tracker &$0.971^{+0.015}_{-0.0093} $&$ -0.11^{+0.33}_{-0.86}$ & $0.965^{+0.014}_{-0.012}$ & $-0.53^{+0.14}_{-0.44}$\\
\Xhline{3\arrayrulewidth}
&\textbf{Union3}& & \hspace{0.5in}   \textbf{Union3+BAO}& \\
\Xhline{3\arrayrulewidth}
CPL &$-0.69\pm 0.15 $& $-0.85\pm 0.79$ & $-0.746\pm 0.094$ & $-0.40^{+0.43}_{-0.35}$\\
BA & $-0.73\pm 0.12$ &$ -0.96 \pm 97$ & $-0.768\pm 0.080 $& $-0.43^{+0.49}_{-0.39}$ \\
JBP &$-0.64^{+0.20}_{-0.14}$ & $-1.4^{+1.0}_{-1.4}$ & $-0.70\pm 0.14$ & $-0.81\pm 0.83$\\
SCPL &$-0.836\pm 0.063$ & $-0.0 \pm 2.3$ & $-0.823\pm 0.055$ & $-0.6^{+1.2}_{-3.3}$\\
GCG-thawer &$0.731 \pm 0.098$ & $-2.18^{+0.29}_{-0.81}$ & $0.724 \pm 090$ & $-2.05^{+0.54}_{-0.85}$\\
GCG-tracker &$0.882^{+0.071}_{-0.049}$ & $-0.41^{+0.15}_{-0.58}$ & $0.856 \pm 0.056$ & $-0.767^{+0.050}_{-0.22}$\\
\Xhline{3\arrayrulewidth}

\end{tabular} 
\label{tab:all model}
\end{table}

\section{Conclusions}\label{sec:conclude} 
In this section, we conclude our analysis. Recent observations have revealed significant deviations (over $2\sigma$) from the standard $\Lambda$CDM model \citet{DES:2024zpp, DESY5}. Furthermore, tensions within the model have become more apparent in recent years, particularly when measuring various cosmological parameters across different observational scenarios. Most famous tensions include the tensions surrounding the Hubble constant $H_0$ (\citet{Verde:2019ivm, Riess:2019qba, DiValentino:2020zio, DiValentino:2021izs, Perivolaropoulos:2021jda, Schoneberg:2021qvd, Shah:2021onj, Abdalla:2022yfr, DiValentino:2022fjm, Hu:2023jqc} and the amplitude of matter fluctuations $S_8$ 
 (\citet{DiValentino:2018gcu, DiValentino:2020vvd, Nunes:2021ipq}). These discrepancies indicate the potential presence of new physics within the Dark Energy sector, highlighting the need for a deeper exploration of alternative models and theoretical frameworks to achieve a more precise cosmological understanding \citet{Eleonora2021, Kamionkowski2022, Sunny2023}. In this work, we have considered six distinct dark energy models and have studied them using recent cosmological observations. For all the dark energy models using different cosmological dataset, We find a significant deviation from $\Lambda$CDM model of more than $2\sigma$ order as shown in subsection \ref{subsec:comparisons}. Additionally, in Section \ref{sec:results}, we showed that the constraints for all dark energy models were most tighter when using the LSST data set, due to the large number of data points available from LSST observations. However, it is important to note that only with the LSST and BAO compilations, the constraints did not improve much, indicating no significant effect of BAO data on LSST observations. This was not observed with other data sets such as Pantheon-Plus, DESY5, and Union3, where the constraints improved with the addition of BAO compilations. Exploring these models with the latest DESI BAO compilations \citet{DESI2024} could reveal improvements over the BAO 2021 compilations. In our work, we did not find good constraints on the $w_a$ parameter for the SCPL and GCG models using local observational data sets. In the future, we plan to incorporate early-time observations, such as those from Planck, to further explore these models in the context of recent cosmological tensions. Additionally, the LSST survey, once operational from 2025 onwards, will release data over the next ten years. While this work is only based on three years of simulated LSST SNIa data, the richer data provided by LSST will help address the recent emerging tensions and shed more light on the need to explore models beyond the $\Lambda$CDM model with greater precision.

\section{Data Availability}

The data used in this work is available from authors upon request.

\acknowledgments 
AM thanks  Richard Kessler and the LSST DESC TD team for the production of the LSST dataset. AAS acknowledges the funding from SERB, Govt of India under the research grant no. CRG/2020/004347. DK acknowledges the CTP, JMI for the research facilities, and MNIT Jaipur for the computing facilities.

\medskip

\bibliographystyle{unsrtnat}
\bibliography{bibliography}
\end{document}